\documentclass[conference]{IEEEtran}
\usepackage{array}
\usepackage{booktabs} 
\usepackage{multirow}
\usepackage{lipsum}
\usepackage{algorithm}
\usepackage{color, soul}
\usepackage{amsmath}
\usepackage{graphicx}
\usepackage{mdframed}
\usepackage{url}
\usepackage{cite}
\usepackage[noend]{algpseudocode}
\usepackage[utf8]{inputenc}
\usepackage{comment}
\usepackage{hyperref}

\newcolumntype{P}[1]{>{\centering\arraybackslash}m{#1}}

\ifCLASSINFOpdf
\else
\fi
\begin{document}
%
\title{Domain Knowledge in Requirements Engineering: A Systematic Mapping Study}

\author{\IEEEauthorblockN{Marina Araújo, Júlia Araújo, Romeu Oliveira, Lucas Romao, Marcos Kalinowski}
\IEEEauthorblockA{Department of Informatics\\
Pontifical Catholic University of Rio de Janeiro (PUC-Rio)\\
Rio de Janeiro, Brazil\\
Emails:\{maraujo, jcaraujo, rferreira, lromao, kalinowski\}@inf.puc-rio.br}
}

%


\maketitle

\begin{abstract}
[Context] Domain knowledge is recognized as a key component for the success of Requirements Engineering (RE), as it provides the conceptual support needed to understand the system context, ensure alignment with stakeholder needs, and reduce ambiguity in requirements specification. Despite its relevance, the scientific literature still lacks a systematic consolidation of how domain knowledge can be effectively used and operationalized in RE. [Goal] This paper addresses this gap by offering a comprehensive overview of existing contributions, including methods, techniques, and tools to incorporate domain knowledge into RE practices. [Method] We conducted a systematic mapping study using a hybrid search strategy that combines database searches with iterative backward and forward snowballing. [Results] In total, we found 75 papers that met our inclusion criteria. The analysis highlights the main types of requirements addressed, the most frequently considered quality attributes, and recurring challenges in the formalization, acquisition, and long-term maintenance of domain knowledge. The results provide support for researchers and practitioners in identifying established approaches and unresolved issues. The study also outlines promising directions for future research, emphasizing the development of scalable, automated, and sustainable solutions to integrate domain knowledge into RE processes. [Conclusion] The study contributes by providing a comprehensive overview that helps to build a conceptual and methodological foundation for knowledge-driven requirements engineering.

\end{abstract}


%

\section{Introduction} \label{sec:introduction}

Domain knowledge plays a fundamental role in Requirements Engineering (RE), as it provides the context for understanding, eliciting, and validating requirements~\cite{zave1997four}. In the context of software development, domain knowledge refers to the specialized understanding of the field in which a system operates. This knowledge is essential to ensure that the software meets the user's needs and is aligned with their objectives. However, capturing and integrating domain knowledge into the requirements process presents significant challenges, particularly in complex and evolving domains~\cite{cao2008agile}, given that domain knowledge is often tacit, residing in human expertise and embedded within organizational processes~\cite{ryan2013acquiring}.


The relation between domain knowledge and requirements engineering (RE) has been extensively studied. Studies have demonstrated that domain knowledge enhances the quality of requirements by reducing ambiguity and increasing specificity. In this context, research indicates that understanding the domain is as crucial as understanding the technical aspects of software development. Without adequate domain knowledge, RE activities risk becoming superficial, leading to requirements that fail to fully address stakeholder needs~\cite{niknafs2012impact}.

Over the years, several solution proposals to enhance domain knowledge integration emerged in RE. Techniques such as knowledge-based requirements engineering, domain ontologies, and model-driven approaches have been proposed to formalize and structure domain knowledge within the RE process~\cite{umar2025automated}. Additionally, artificial intelligence (AI) and machine learning (ML) are being explored as tools for extracting domain knowledge from existing data sources~\cite{li2024machine}.

Recognizing the importance of domain knowledge in RE, we contribute a systematic analysis of existing research on this topic. Our objective is to characterize the role of domain knowledge in RE in terms of its impact on requirements quality, challenges in capturing and utilizing it, and the methodologies employed to enhance domain knowledge integration in RE processes. Therefore, we conducted a systematic mapping study, strictly following the search strategy recommended by Wohlin \textit{et al.}~\cite{wohlin2022successful}. We identified 75 articles that present contributions related to domain knowledge to support RE.

Our findings indicate that several studies have applied domain knowledge-based solutions to support key RE activities, such as requirements elicitation, analysis, and validation. Furthermore, we identify key challenges such as knowledge transfer between domain experts and software engineers, evolving domain landscapes, and the need for structured knowledge representation. Addressing these challenges is essential for ensuring that RE practices effectively leverage domain knowledge to produce high-quality software systems.

The remainder of this paper is structured as follows. Section~\ref{sec:background} outlines the background and provides an overview of the related work. Section~\ref{sec:protocol} details the mapping study protocol and its implementation. Section~\ref{sec:results} presents the results of the mapping study. Section~\ref{sec:conclusions} discusses the findings. Section~\ref{sec:threats} discusses potential threats to the validity of our study. Lastly, Section~\ref{sec:remarks} offers the concluding remarks.

\section{Background and Related Work} \label{sec:background}

Effective RE is fundamentally intertwined with a deep understanding of the problem domain. Domain knowledge, encompassing the concepts, relationships, and constraints specific to a particular field, is crucial for eliciting, analyzing, specifying, and validating requirements that are both accurate and relevant~\cite{zave1997four}. This section lays the background for a deeper understanding of the role and management of domain knowledge within the RE process.

An important aspect of RE is the need for stakeholders to have sufficient expertise to effectively communicate their needs and expectations. Without a solid grasp of the domain, stakeholders may have difficulty articulating requirements precisely, leading to ambiguity, incompleteness, and, ultimately, to a system that fails to meet its intended purpose. This is particularly relevant in complex or specialized domains~\cite{aranda2015effect}~\cite{callele2005requirements}~\cite{hadar2014role}~\cite{siqueira2014transforming}~\cite{zave1997four}. Furthermore, requirement validation is highly dependent on domain knowledge. Ensuring that requirements are consistent, complete, and feasible requires a thorough understanding of their constraints, limitations, and potential risks~\cite{machanavajjhala2007diversity}~\cite{sweeney2002k}, as well as the implicit need for feasibility assessment~\cite{chen2023use}.

The selection and application of appropriate RE techniques, as well as effective management of evolving requirements, also depend on domain expertise~\cite{agrawal1993mining}~\cite{guo2018domain}~\cite{kim2019ontology}. It is important that all terminology used in RE be related to the reality of the environment in which the system will be used. Zave and Jackson~\cite{zave1997four} emphasized the importance of avoiding vague or abstract terms, which can lead to erroneous interpretations and ambiguities. To ensure clarity, each term employed in RE should be precisely defined, connecting it to concrete, real-world concepts. The work by Parnas and Madey~\cite{parnas1995functional} and van Schouwen \textit{et al}~\cite{van1993documentation} presents a methodology and a set of tools for documenting software requirements, emphasizing the importance of clear and precise specifications. Their approach highlights the need for a shared understanding of the domain between stakeholders and developers, which is facilitated by the use of structured documentation and formal methods~\cite{aranda2015effect}~\cite{hadar2014role}~\cite{siqueira2014transforming}~\cite{zave1997four}.

The aforementioned studies recognize and reinforce the role of domain knowledge in RE. However, we did not identify research specifically focused on investigating and synthesizing the state of the art of approaches based on the acquisition and management of domain knowledge to support RE activities.

\section{Systematic Mapping Protocol}  \label{sec:protocol}

Systematic Mapping (SM) studies are designed to provide a broad overview of a research area, identify the existence of evidence on a topic, and indicate the quantity of such evidence. The SM study was conducted following the guidelines proposed by Kitchenham and Charters~\cite{keele2007guidelines}, as well as the SM-specific guidelines by Petersen \textit{et al.}.~\cite{petersen2015guidelines}. After identifying the need for the review, we defined the research questions, search strategy, and inclusion/exclusion criteria.

\subsection{Goal and Research Questions}

The main goal of this systematic mapping is to identify and analyze research that leverages domain knowledge to support RE activities. To support this objective, the following research questions were defined.

\textbf{RQ1. What domain knowledge contributions have emerged to support RE?} This question aims to provide an overview of the contributions related to domain knowledge that support the execution of RE activities.

\textbf{RQ2. What types of domain knowledge are addressed by
the identified contributions?} The goal of this question is to identify the different types of domain knowledge considered (tacit, explicit knowledge, or both), helping to understand its role in the elicitation and specification of requirements.

\textbf{RQ3. What types of requirements were most addressed by the domain knowledge-related contributions?} Here, the aim is to investigate whether the identified contributions focus on functional or non-functional requirements.


\textbf{RQ4. What quality attributes or non-functional requirements (NFRs) were most addressed in domain knowledge-related research?} According to the ISO 25010~\cite{iso25010:2023}, domain knowledge can impact different quality characteristics (\textit{e.g.}, performance, maintainability, and usability). This question aims to identify which quality attributes were most addressed.

\textbf{RQ5. What are the main research challenges reported at the intersection of domain knowledge and RE?} This question seeks to identify unresolved issues and obstacles in integrating domain knowledge into RE, highlighting gaps that hinder progress in this field.

\textbf{RQ6. What are the main research directions reported at the intersection of domain knowledge and RE?} This question aims to identify opportunities for future research, helping to guide new studies and investigations.

\textbf{RQ7. What types of research were identified in the contributions?} The goal of this question is to classify the studies according to their research type facets, adopting the taxonomy by Wieringa et al.~\cite{wieringa2006requirements}. This will help understand the nature of the contributions.

\textbf{RQ8. What types of empirical evaluations were conducted to assess the contributions?} This question aims to identify the empirical methods used (experiments, comparative studies, case studies), providing insight into the scientific rigor and maturity of the evidence in the field.

While questions RQ1-RQ4 aim at structuring the publication landscape in a conceptual manner, RQ5 and RQ6 focus on identifying challenges and future research directions, finally RQ7 and RQ8 provide insights into the nature of the current reported evidence.

\subsection{Search Strategy}

The mapping study used a hybrid search strategy~\cite{wohlin2022successful}, which involved performing a database search with a search string on a particular digital library (Scopus) to identify a representative seed set and then applying iterative backward and forward snowballing (using Google Scholar). We chose this hybrid strategy as it has shown to be effective to identify relevant primary studies~\cite{wohlin2022successful}. We selected the more comprehensive iterative snowballing technique to maximize recall, even though it would mean analyzing a greater number of papers~\cite{mourao2020performance}. Iterative backward and forward snowballing entails applying both backward and forward snowballing on every newly included paper. 

To begin the database search on Scopus, we crafted the search string using the PICO (Population, Intervention, Comparison, Outcome) criteria~\cite{leonardo2018pico}. Our study focuses on RE (population) and investigates the contribution of domain knowledge (intervention) to this field. Since this is a mapping study, there was no need for specific comparisons or limiting the outcomes, so we only required keywords related to RE and domain knowledge. The defined search string, to be applied on titles, abstracts and keywords was: \textit{“Requirements Engineering” AND “Domain Knowledge”}. It is noteworthy that this search string is only used to retrieve and filter a representative initial seed set, which is then complemented by applying detailed iterative backward and forward snowballing procedures. 

\subsection{Study Selection}

The primary inclusion criterion focused on studies presenting contributions related to domain knowledge that support RE (IC1). If multiple papers reported the same study, only the most recent one was included.

To ensure the quality and relevance of the selected papers, we applied specific exclusion criteria. RE is a research area with a high volume of publications and numerous contributions from different perspectives. Given this extensive body of work, it was essential to apply a filtering strategy to focus on the most impactful studies. Therefore, we restricted our selection to papers published in the top 20 venues in software engineering (SE) and the top 5 venues in RE, as ranked by Google Scholar. This approach ensured that our dataset consisted only of rigorously peer-reviewed research from the most authoritative sources while keeping the study manageable. Additionally, we excluded papers that do not include, in the title, abstract, or keywords, at least the terms `Requirements' or `Domain', in order to retain only studies with a direct connection to the topic under investigation. Finally we excluded grey literature and short papers. The selection criteria applied for filtering the papers are shown in Table~\ref{tab:selectioncriteria}. 

\begin{table}[h]
\centering
\scriptsize
\caption{Selection criteria.}
\label{tab:selectioncriteria}
\begin{tabular}{|c|p{6,6cm}|}
\hline
\textbf{Criteria} & \multicolumn{1}{|c|}{\textbf{Description}}\\ 
\hline
IC1  & Articles that present contributions related to domain knowledge that have emerged to support RE.\\ \hline
EC1  & Papers that are not full papers published in the main venues of the SE field or in the leading venues specifically focused on RE. \\ \hline
EC2  & 
Papers that not include, in the title, abstract, or keywords, terms such as 'Requirements' or 'Domain'\\ \hline
EC3  & 
Theses and dissertations, book chapters, calls for papers, basic teaching materials, and short papers (less than 6 pages).\\ \hline

\hline
\end{tabular}
\end{table}

The selected SE venues were the Google Scholar top-ranked venues in the area at the time of writing this article\footnote{Google Scholar - Category Software Systems \url{https://scholar.google.com/citations?view_op=top_venues&hl=en&vq=eng_softwaresystems}}: \textit{International Conference on Software Engineering}; \textit{IEEE Transactions on Software Engineering}; \textit{Journal of Systems and Software}; \textit{Proceedings of the ACM on Programming Languages}; \textit{Information and Software Technology}; \textit{Empirical Software Engineering}; \textit{International Symposium on Foundations of Software Engineering}; \textit{International Conference on Automated Software Engineering}, \textit{ACM SIGPLAN Conference on Programming Language Design and Implementation}; \textit{ACM Transactions on Software Engineering and Methodology}; \textit{Mining Software Repositories}; \textit{International Symposium on Software Testing and Analysis}; \textit{IEEE Software}; \textit{Software: Practice and Experience}; \textit{Software and Systems Modeling}; \textit{Symposium on Operating Systems Principles}; \textit{International Conference on Software Analysis, Evolution, and Reengineering}; \textit{International Conference on Tools and Algorithms for the Construction and Analysis of Systems}; \textit{Symposium on Principles and Practice of Parallel Programming}; and the \textit{International Conference on Software Maintenance and Evolution}.

For RE, we specifically included the \textit{IEEE International Requirements Engineering Conference (RE Conference)}, the \textit{Requirements Engineering Journal (RE Journal)}, the \textit{International Working Conference on Requirements Engineering: Foundation for Software Quality (REFSQ)}, the \textit{International Conference on Advanced Information Systems Engineering (CAiSE)}, and \textit{ACM Symposium on Applied Computing – Requirements Engineering Track (ACM SAC - RE Track)}. CAiSE and ACM SAC were included because they have a dedicated RE track, making them relevant for research in the field. 

Given that the selection criteria already imposed strict restrictions regarding the source of the publications (only full papers from high-quality journals and conferences), we assume methodological quality and scientific relevance in the included studies. Therefore, no additional quality assessment criteria were applied in this review. 

\begin{figure}[h]
    \centering
    \includegraphics[width=0.45\textwidth]{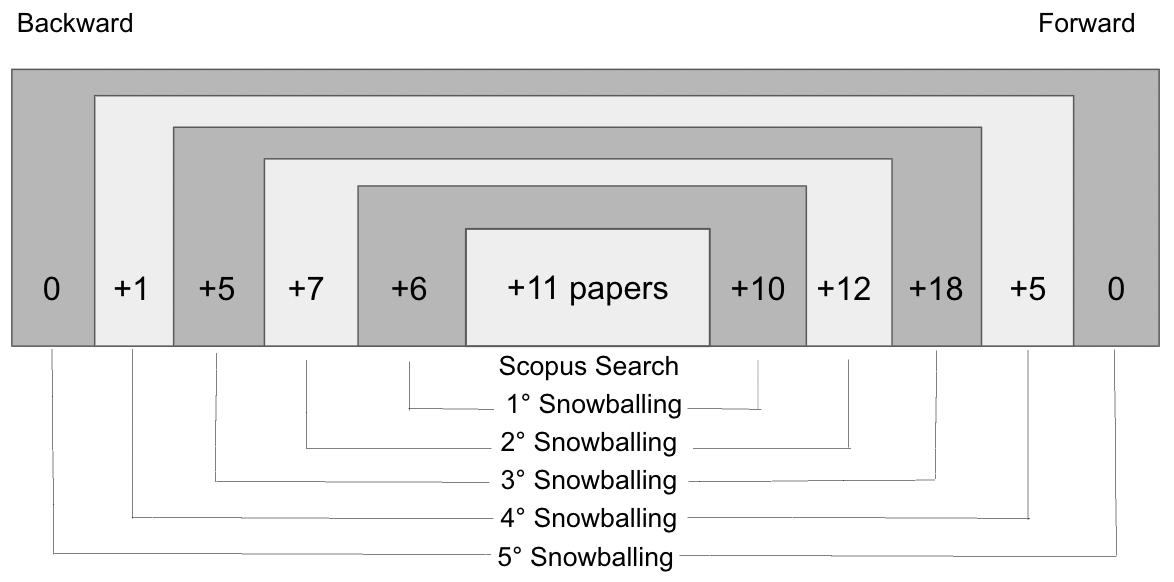}
    \caption{Papers selection process.}
    \label{fig:bfsnowballing}
\end{figure}

Fig.~\ref{fig:bfsnowballing} illustrates the steps carried out in the paper selection process. The first step consisted of conducting a search using the search string in the digital library selected for this study. The search string was applied to titles, abstracts, and keywords in Scopus, in February 2025, resulting in 289 papers. Exclusion criterion EC1 was then applied, removing papers that were not published in the selected venues, reducing the set to 66 papers. Next, exclusion criterion EC2 was applied, eliminating all papers that did not contain the terms `domain' or `requirements' in the title, abstract, or keywords. Finally, EC3 excluded grey literature and short papers. As a result, 11 papers were included.

In the following stage, carried out during the months of March and April, the iterative backward and forward snowballing process was applied, following the guidelines established by~\cite{wohlin2014guidelines}. A total of five backward snowballing (BS) and five forward snowballing (FS) iterations were conducted until the final set of studies was consolidated (\textit{i.e.}, no more cited or citing studies to be included were found). The ten snowballing iterations (BS and FS) resulted in analyzing 1,687 papers that passed the exclusion criterion EC1 (\textit{i.e.}, that were published in the selected venues). 

After removing duplicates, applying the inclusion criterion and the exclusion criteria EC2 and EC3, 64 papers were selected to compose the final set obtained through snowballing. Therefore, the study included a total of 75 papers: 11 from the Scopus search and 64 from the snowballing process. The detailed filtering process, including all papers analyzed from Scopus and from each snowballing iteration, as well as the application of the criteria to each paper, is transparently available in our open science repository~\cite{zenodoRepository}.

\subsection{Data Extraction and Classification Scheme}
The data and information collected from the selected papers, along with the classification schemes outlining the various categories, are presented in  Table~\ref{tab:dataExtraction}. The full set of extracted data is also accessible in our online open science repository~\cite{zenodoRepository}.

\begin{table}[h]
\centering
\scriptsize
\caption{Data Extraction Form.}
\label{tab:dataExtraction}
\begin{tabular}{|p{2cm}|p{6cm}|}
\hline
\multicolumn{1}{|c|}{\textbf{Information}} & \multicolumn{1}{c|}{\textbf{Description}}\\ 
\hline
Domain Knowledge Contributions (RQ1) & Contributions related to domain knowledge that emerged to support RE.\\ \hline
Types of Domain Knowledge (RQ2)  & Types of domain knowledge that were addressed by the identified contributions.\\\hline
Requirement Types (RQ3) & Types of requirements that were most addressed by the contributions related to domain knowledge. \\\hline
Quality Attributes / Non-Functional Requirements (RQ4) & Quality attributes or non-functional requirements (NFRs) that were most addressed in research related to domain knowledge.\\ \hline
Research Challenges (RQ5) & Major research challenges reported at the intersection of domain knowledge and RE.\\\hline
Research Directions (RQ6)  & Main research directions reported at the intersection of domain knowledge and RE.\\\hline
Research Type Facet (RQ7) & The classification of research types: evaluation research, solution proposal, philosophical paper, opinion paper, or experience paper, as outlined by Wieringa \textit{et al.}~\cite{wieringa2006requirements} \\\hline
Empirical Evaluation (RQ8) & Classification of the empirical strategy~\cite{wohlin2024experimentation}, including categories such as experiment and case study.\\\hline

\hline
\end{tabular}
\end{table}
\section{Systematic Mapping Results}  \label{sec:results}

This section presents the results of the systematic mapping study on Domain Knowledge for RE. The selected studies were analyzed based on predefined classification criteria, aiming to provide a structured overview of how domain knowledge has been addressed in the context of RE. Due to space constraints, we had to move the references of the 75 included studies (P1-P75) to an online supplement in our open science repository~\cite{zenodoRepository}. The results are detailed hereafter.

\subsection{RQ.1 What domain knowledge contributions have emerged to support RE?}

The literature review indicates that contributions related to the use of domain knowledge in RE are predominantly grouped into three categories: methods, techniques, and tools, as presented in Table~\ref{tab:contributions}.

\begin{table}[h]
\centering
\scriptsize
\caption{Identified Contributions.}
\label{tab:contributions}
\begin{tabular}{|p{1 cm}|p{3 cm}|p{3.5 cm}|}
\hline
\multicolumn{1}{|c}{\textbf{Category}} & \multicolumn{1}{|c|}{\textbf{ID}} & \multicolumn{1}{c|}{\textbf{Examples}} \\ 
\hline
Method (3) & P3, P11, P69 & Some studies propose structured methods for using domain knowledge. One transforms textual requirements into production rules [P3], another uses conceptual modeling for complex domains [P11], and a third applies logic-based structures to formalize semantics [P69].\\\hline
Method and Technique (11) & P1, P10, P14, P16, P17, P19, P21, P22, P57, P61, P68 & Several works combine formal modeling with semantic techniques. Some align models with domain terms [P1], reuse prior requirements via similarity [P10], or integrate automated extraction with structured modeling [P14, P16, P21, P22]. \\\hline
Method and Tool (4) & P35, P38, P39, P59 & Some methods are implemented with tool support. Examples include semantic visualizations [P35], interactive categorization [P38, P39] and automated validation [P59].\\ \hline
Technique and Tool (7) & P4, P25, P36, P37, P42, P65, P67 & These studies merge techniques like clustering and NLP with tools for elicitation and analysis. Applications include collaborative modeling [P4], similarity detection [P25, P36], and semantic mapping [P42, P67].\\\hline
Method, Technique and Tool (46) & P2, P5, P6, P7, P9, P12, P13, P15, P18, P20, P23, P24, P26, P27, P28, P29, P30, P32, P33, P34, P40, P41, P44, P45, P46, P47, P49, P50, P51, P52, P53, P54, P55, P56, P58, P60, P62, P63, P64, P66, P70, P71, P72, P73, P74, P75 & This is the most common category. Examples include Domain Theory combining ontologies and tools [P2], automated extraction using embeddings [P5], semantic reasoning in critical systems [P6], and hybrid frameworks for intelligent elicitation [P13]. Many others follow similar patterns.

\\\hline
Other (4) & P8, P31, P43, P48
& Some works offer theoretical reflections, methodologies and frameworks. One discusses tacit knowledge challenges [P8], another highlights gaps in formalization [P43], and a third critiques traditional elicitation in agile settings [P48].

\\\hline
\end{tabular}
\end{table}

These contributions aim to formalize and operationalize knowledge derived from experts or structured sources, with the objective of enhancing key processes such as requirements elicitation, analysis, and validation. The category analysis reveals that 3 articles were identified in the exclusive method category, the method and technique combination includes 11 articles, method and tool accounts for 4 articles, technique and tool includes 7 articles and the most comprehensive combination, method, technique, and tool, comprises 46 articles. Finally, the other category, which includes methodology, framework, and evaluation, encompasses 4 articles. Notably, no study was classified solely as a technique or a tool without being integrated with other approaches. 


\subsection{RQ.2 What types of domain knowledge are addressed by the identified contributions?}
The vast majority of the selected papers address explicit domain knowledge (68 out of 75), emphasizing formalization, representation, and computational manipulation. A smaller number of contributions explore tacit knowledge (2 papers), often related to expert judgment or experiential insights. Additionally, 5 studies propose hybrid approaches, combining both explicit and tacit elements to enrich the elicitation and modeling of requirements. These results highlight a predominant focus on knowledge that can be structured and encoded, with limited attention to more experiential or informal dimensions.


\subsection{RQ.3 What types of requirements were most addressed
by the domain knowledge-related contributions? }

An analysis of the selected studies reveals distinct emphases regarding the types of requirements addressed through domain knowledge. Among the 75 contributions reviewed, the majority focused on functional requirements (34 papers), indicating a prevailing concern with specifying system behaviors and functionalities grounded in domain expertise. In addition, 24 studies addressed non-functional requirements, and 17 contributions considered both types simultaneously. This distribution highlights the predominant attention given to functional aspects, while also evidencing a meaningful, though less frequent, effort to use domain knowledge to address quality-related attributes in the RE process.

\subsection{RQ.4 What quality attributes were most addressed in domain knowledge-related research? }

The analysis of the contributions enabled the identification of a variety of quality requirements, also referred to as non-functional requirements (NFRs), addressed in the selected studies. Table~\ref{tab:quality_characteristics} presents the frequency with which these quality characteristics are considered across the papers. The definition and categorization of these quality criteria are based on the ISO 25010~\cite{iso25010:2023}.It is possible to observe that domain knowledge has been mainly considered to support reliability, compliance, usability, and security-related issues.

\begin{table}[h]
\centering
\scriptsize
\caption{Frequency of quality characteristics.}
\label{tab:quality_characteristics}
\begin{tabular}{|l|c|l|c|}
\hline
\multicolumn{1}{|l|}{\textbf{Characteristic}} & \multicolumn{1}{l|}{\textbf{Frequency}} & \multicolumn{1}{l|}{\textbf{Characteristic}} & \multicolumn{1}{l|}{\textbf{Frequency}}\\
\hline
Adaptability  & 5 & Usability & 23 \\\hline
Portability  & 0 & Compatibility & 3  \\\hline
Security  & 20 & Performance & 8 \\\hline
Reliability  & 28 & Maintainability & 9 \\\hline
Compliance  & 23 & \multicolumn{1}{c|}{-}  & - \\\hline
\end{tabular}
\end{table}

\subsection{RQ.5 What are the main research challenges reported
at the intersection of domain knowledge and RE?}
The integration of domain knowledge into RE presents a range of challenges, which have been consistently reported across the literature. Based on the analysis of the 75 selected papers, the challenges can be grouped into three themes: technical, acquisition, and management challenges. 

\textbf{Technical challenges}: This was the most frequently reported category in the analysis. Common issues include the complexity of integrating domain knowledge into modeling and reasoning tools [P2, P5, P12], limitations in the scalability of formal representations [P1, P27, P33], and the lack of tool support for semantic or rule-based processing [P3, P6, P13, P42, P46]. Other studies pointed to the need for advanced mechanisms to maintain traceability and consistency during the application of domain knowledge [P36, P44, P52, P64].

\textbf{Acquisition challenges}: Refer to the difficulties in capturing, formalizing, and structuring domain knowledge. Many papers highlighted the lack of accessible documentation or expert availability, especially in complex or regulated domains [P7, P9, P14, P28, P53]. Others emphasized the need for automated or semi-automated extraction methods [P26, P34, P38, P51, P61], as well as barriers to generalizing tacit knowledge into reusable forms [P13, P25, P29].

\textbf{Management challenges}: This was the least reported category, these challenges revolve around the governance, evolution, and standardization of domain knowledge over time. For instance, papers such as P10, P16, and
P20 mentioned difficulties in maintaining knowledge consistency across system updates or organizational changes. Others [P22, P27, P31, P54] discussed the lack of versioning strategies, reuse policies, and the need for institutional frameworks to support the long-term management of knowledge assets.

\subsection{RQ.6 What are the main research directions reported
at the intersection of domain knowledge and RE?}

The analysis of research directions reported in the selected articles reveals a consensus on the need for continuous advancements on multiple fronts, which can be grouped into seven themes.

First, \textbf{Enhancement of Techniques and Methods} proposes the development and refinement using AI approaches through the application of advancements in Natural Language Processing (NLP), Large Language Models (LLMs), and machine learning (ML) [P5, P6, P32, P50]. Future research seeks to improve the extraction, representation (\textit{e.g.}, via ontologies and knowledge graphs [P2, P29]), and reasoning about domain knowledge to identify, analyze, and validate requirements more accurately, addressing challenges such as ambiguity [P7, P71], variability [P35, P43], and tacit requirements [P35]. Increasing the explainability and interpretability of applied AI models is also a concern [P49, P59].

Second, \textbf{Generalization, Scalability, and Applicability} point to the need to adapt and validate techniques new domains (e.g., healthcare, finance, building and infrastructure industry, critical systems) [P3, P14, P15], deal with different languages and requirements formats [P16, P73], and ensure that approaches are scalable for large-scale systems [P9, P28].

Third, \textbf{Automation and Tool Support} aims to reduce manual effort and increase the efficiency of requirements engineers. Directions include the development of more sophisticated tools to partially or fully automate the construction of domain models [P5, P12], the extraction and annotation of requirements [P13, P42], the generation of artifacts [P39, P53], and integration with development environments [P2, P23].

Fourth, \textbf{User Focus, Collaboration, and Interaction} recognizes the importance of the human factor. Future research aims to improve the involvement of experts and stakeholders in the RE lifecycle [P1, P31], develop mechanisms for collecting and incorporating user feedback [P24, P26, P36], support negotiation and collaborative conflict resolution [P28, P69], and explore the personalization of RE approaches [P57].

Fifth, \textbf{Integration and Interoperability} seeks creating more holistic solutions. This includes the integration of different logics and modeling paradigms [P1,
P33], the combination of symbolic and learning-based techniques [P52], and a stronger connection with other SE artifacts and processes, such as architectural design and defect analysis [P35, P45, P56].

Sixth \textbf{Validation and Empirical Evaluation} future directions include conducting more case studies in real industrial contexts [P3, P61], controlled experiments, usability studies with requirements engineers [P6, P9], and the development of benchmarks and metrics for comparative evaluation of approaches [P19, P59].

Finally, \textbf{Knowledge Management, Evolution, and Quality} are important directions to ensure the long-term relevance and utility of domain knowledge in RE. This involves research on how to maintain and evolve domain models and ontologies [P8, P21, P38], ensure traceability and consistency of requirements [P31, P45], manage volatility, and identify missing or inconsistent requirements [P8, P51, P75].

\subsection{RQ.7 What types of research were identified in the
contributions?}
To structure the data regarding the type of research in the analyzed papers, we initially considered the primary intent of each study. For example, if a given paper proposed a method, technique, or tool, it was classified as a “Solution Proposal”, even if it included some empirical study. In such cases, the evaluation or validation of the proposal was regarded as a secondary intent, serving as methodological support to empirically substantiate the proposed approach. By considering this, most of the analyzed papers were classified as “Solution Proposal”, totaling 69 studies. Research with a primary intent related to evaluation accounted for 5 papers. Additionally, 1 paper was identified as an “Opinion Paper”. Considering the primary focus, no validation research, philosophical papers, or personal experience papers were identified. Table ~\ref{tab:identifiedcontibutions} presents a quantitative summary of the articles grouped by research type.

\begin{table}[h]
\centering
\scriptsize
\caption{Research Types.}
\label{tab:identifiedcontibutions}
\begin{tabular}{|p{2cm}|p{6cm}|}
\hline
\multicolumn{1}{|c|}{\textbf{Information}} & \multicolumn{1}{c|}{\textbf{Description}}\\ 
\hline
Research type  & Highlights/ Examples \\\hline
Solution proposal (69) & Development of new methods, techniques, or tools, often accompanied by some form of empirical study. For example, in paper P2, the authors propose a solution evaluated with the support of RE experts, including cognitive studies (such as card sorting) and application in real-world scenarios. In P5, the proposal is scenario-based and supported by a tool designed to collect software requirements specific to the building and infrastructure industry. In paper P7, the authors present a method for detecting and interpreting syntactic ambiguity, which is rigorously evaluated using quantitative metrics (precision, recall, and accuracy), comparisons with the Stanford Parser and a generic corpus, as well as validation across seven distinct industrial domains.\\ \hline
Evaluation research (5)  & Evaluations conducted in case studies or industrial settings. Some conducted evaluations were not necessarily empirical. For instance, P8 involved the support of experts for the application and evaluation of the card sorting method as a knowledge elicitation technique in requirements. Additionally, P31 evaluates a methodology for analyzing security and privacy requirements regarding compliance with legislation (HIPAA), using a case study that involves mapping, refinement, and traceability of existing requirements.\\\hline
Opinion paper (1) & Literature review and theoretical development of a conceptual framework. There is no proposal for a technical solution or systematic empirical evaluation, but rather a structured reflection on existing approaches and future directions [P43]. \\\hline

\hline
\end{tabular}
\end{table}

\begin{figure}[h]
    \centering
    \includegraphics[width=0.5\textwidth]{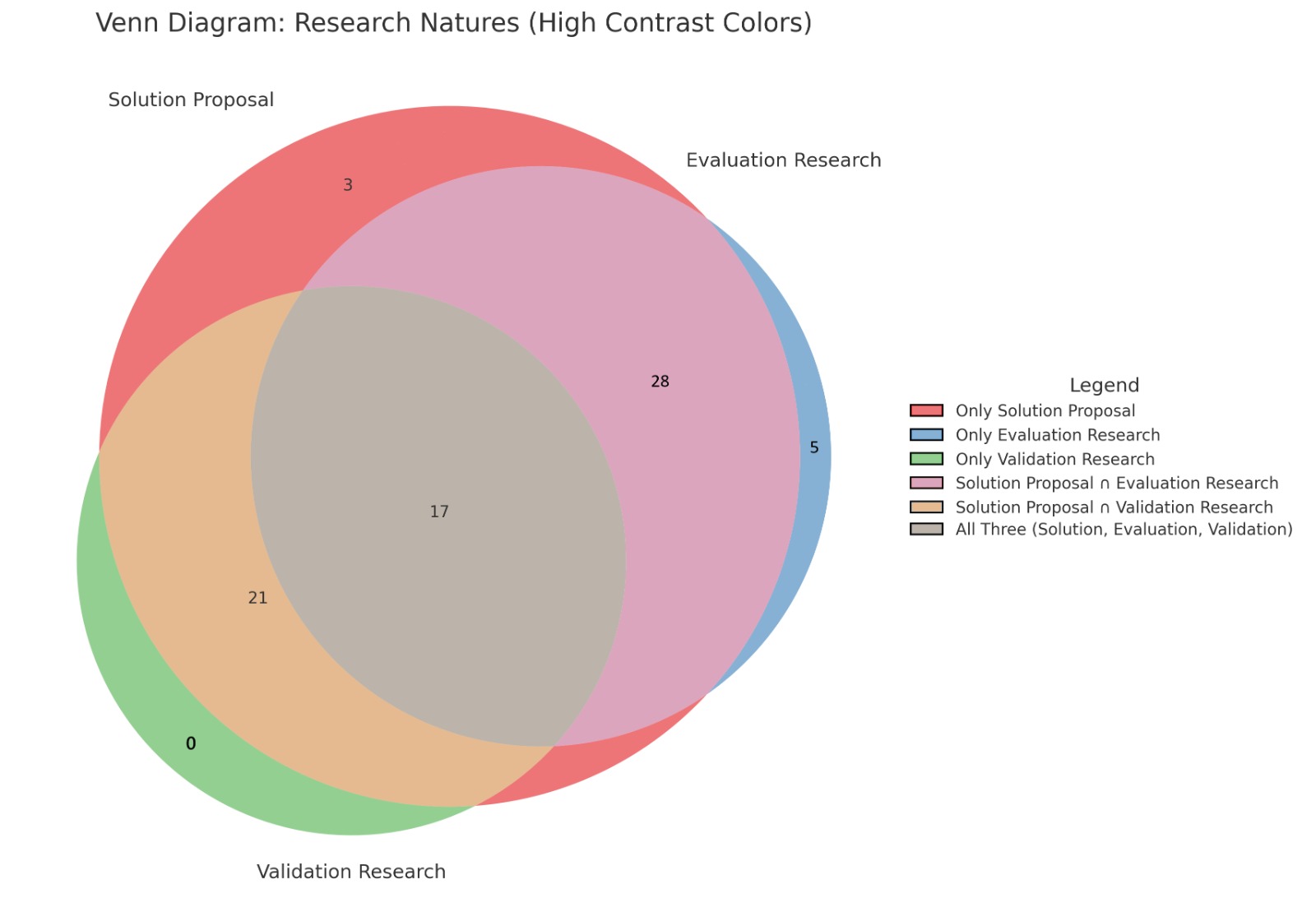}
    \caption{Quantitative summary of the types of research identified and their intersections.}
    \label{fig:nrq7}
\end{figure}

We also highlight the strong intersections between solution proposals (primary intent) and some form of empirical study, as many papers proposed a strategy and have already undertaken some form of evaluation and/or validation. As shown in Fig. ~\ref{fig:nrq7}, only three papers proposed a solution without conducting empirical studies for its evaluation and/or validation. Conversely, 17 papers performed both empirical evaluations and validations of the proposed solution. In the responses to RQ8, we provide a more detailed discussion regarding the types of empirical studies applied.

\subsection{RQ.8 What types of empirical evaluations were conducted to assess the contributions?}

\begin{figure}[h]
    \centering
    \includegraphics[width=0.5\textwidth]{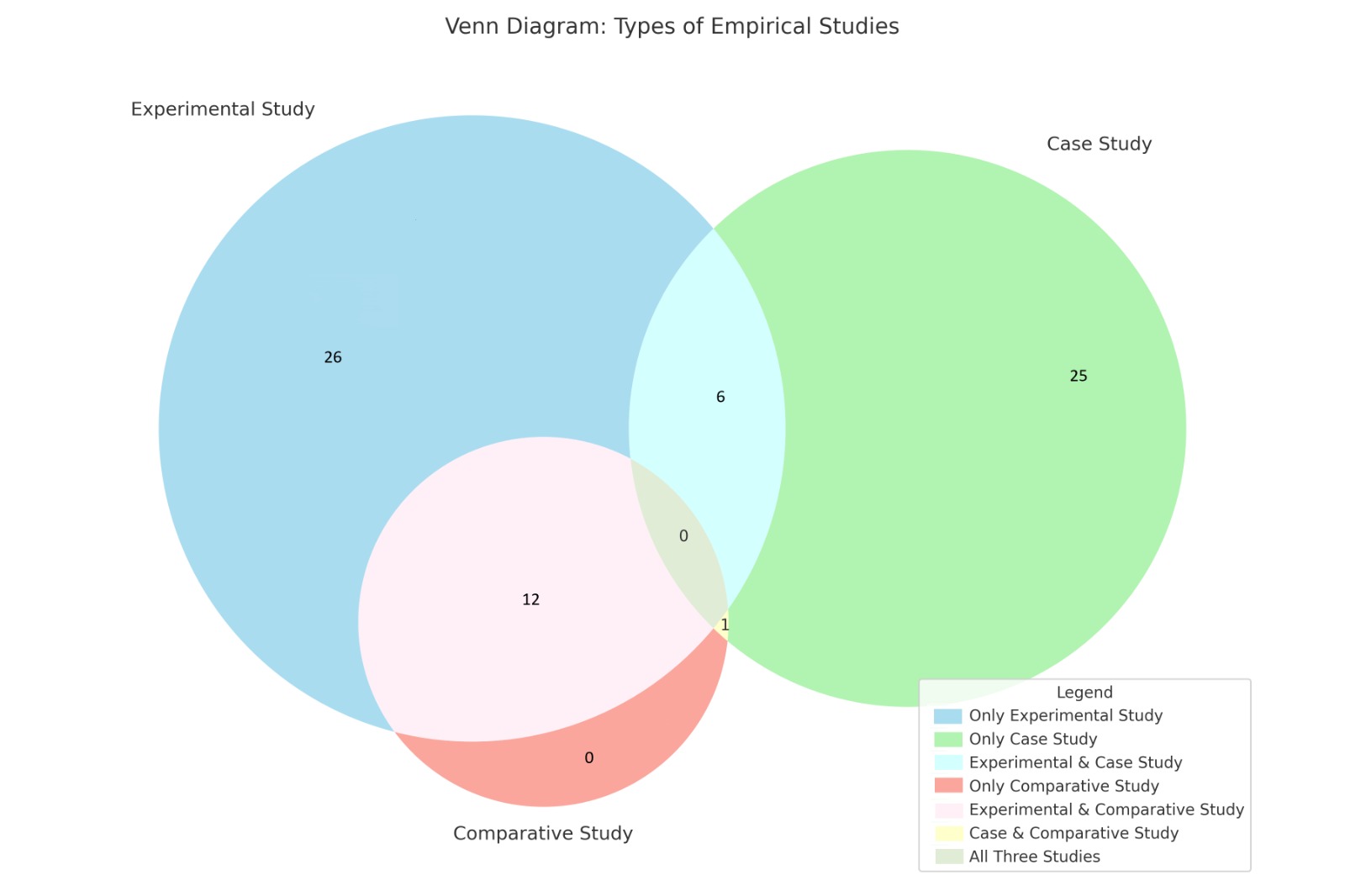}
    \caption{Quantitative summary of the types of empirical studies identified and their intersections.}
    \label{fig:rq8}
\end{figure}

Among the 75 papers included in this mapping, the majority conducted some type of empirical study, totaling 70 papers. It is also worth noting that several of these articles applied more than one type of empirical study (Fig.~\ref{fig:rq8}). For example, some works combined case studies and comparative studies in their research. It was possible to observe that a portion of the studies empirically supported their approaches using only experimental studies, totaling 26 papers, while 25 papers used only case studies. Finally, only 5 papers did not include any type of empirical study.

\section{DISCUSSION}  \label{sec:conclusions}


The results indicate a significant amount of research that involves the application of domain knowledge in RE, with an emphasis on approaches that integrate methods, techniques, and tools. This trend reflects a pursuit of more structured and automated solutions aligned with practical needs. 

We also observed that there is a predominance of approaches focused on explicit knowledge, to the detriment of tacit knowledge, which is essential in complex domains that rely heavily on practical experience. The lack of effective strategies to capture and integrate this type of knowledge reveals an important gap and highlights the need for more robust hybrid solutions capable of articulating different forms of knowledge.

Although non-functional requirements (NFRs) such as reliability, security, and usability are critical for software quality, especially in regulated or high-risk environments, this study reveals a predominant focus on functional requirements. Among the 75 analyzed papers, 34 addressed only functional requirements, compared to 24 that focused on NFRs and 17 that considered both. This imbalance reflects a common issue in software engineering, where NFRs are often postponed or neglected, a practice that can lead to rework, financial loss, and compliance failures~\cite{viviani2023empirical}. 

On the other hand, technical challenges remain, such as difficulties in scaling formal representations, integrating tools, and efficiently maintaining traceability. In addition, barriers persist in knowledge acquisition, whether due to the lack of proper documentation or limited access to experts. Management aspects, such as updating and reusing knowledge over time, are also still underexplored.

Future research directions indicate increasing interest in the use of artificial intelligence, machine learning, and natural language processing to support the extraction and application of domain knowledge. There is also a growing emphasis on expert collaboration and the adaptation of approaches to different organizational contexts.

The predominance of domain knowledge studies classified as solution proposals may reflect a proactive effort by researchers to translate theoretical foundations into practical support for RE activities. However, although most of the analyzed proposals include some form of empirical evaluation, there is still an opportunity to deepen these investigations in industrial settings and with greater methodological rigor. Broader and more replicable studies supported by standardized metrics can play an important role in consolidating existing practices and strengthening the practical impact of the contributions in RE.
\section{THREATS TO VALIDITY}  \label{sec:threats}

This section outlines potential threats to the validity of our systematic mapping study and describes the measures adopted to mitigate them. 

With respect to \textbf{internal validity}, a common threat for systematic mapping studies concerns the application of the search strategy. We designed the search string based on the PICO strategy to ensure direct alignment with our research goals. Furthermore, we employed a hybrid search strategy that combined database searches with iterative backward and forward snowballing. This approach has been recognized as effective in secondary studies, contributing to a representative selection of relevant papers~\cite{wohlin2022successful}. The search strategy was systematically applied and thoroughly documented in our open science repository~\cite{zenodoRepository}, with details on each snowballing iteration and the screening of the papers. 

Regarding \textbf{external validity}, we used a search strategy recognized as a suitable approach for secondary studies. Despite these efforts, it is not possible to completely rule out the possibility that some relevant studies might have been missed. However, no additional publications matching our inclusion criteria were identified through manual searches, which reinforces our confidence in having a representative dataset. The conclusions presented in this work are based exclusively on the evidence reported in the included primary studies. Although all selected studies were peer-reviewed and published in the most influential venues in the field, we did not conduct a formal assessment of methodological quality. This step, while relevant, is generally not part of the scope of mapping studies and may be incorporated in a future systematic review. 

Finally, concerning \textbf{reliability}, to reduce potential bias in study selection, screening, data extraction, and coding were independently conducted by two researchers with peer review. A third researcher mediated any disagreements, which were resolved by consensus, ensuring the consistency and reliability of the results. All methodological details, including the search protocol, data extraction procedures, and coding criteria, are transparently documented and publicly available and auditable in our open science repository~\cite{zenodoRepository}.

\section{CONCLUDING REMARKS}  \label{sec:remarks}

This paper presents the results of a systematic mapping study on domain knowledge-based research aimed at supporting RE activities. We applied a hybrid search strategy, combining a structured search on the Scopus database with iterative backward and forward snowballing using Google Scholar. This approach enabled the identification of a total of 75 primary studies.

We summarized and classified the identified contributions into categories such as methods, techniques, tools, or their combinations. These studies address both functional and non-functional requirements, with particular emphasis on quality attributes like reliability, usability, and security. We also analyzed the types of domain knowledge considered (explicit, tacit, and hybrid), the nature of the research (\textit{e.g.}, solution proposals and evaluation studies), and the empirical strategies adopted for validation. Our analysis revealed recurring technical, acquisition, and management challenges, as well as a wide range of promising research directions. These include the application of AI techniques, improvements in tool support, enhanced user collaboration, and more rigorous empirical validation in industrial contexts.

To the best of our knowledge, this is the first systematic mapping study to consolidate evidence on how domain knowledge has been used to support RE processes. Hence, the main contributions of this study are twofold: (i) providing a structured map of current contributions that integrate domain knowledge into RE practices, a topic of growing importance but that was still lacking systematization; and (ii) identifying critical research gaps and future research opportunities that can inform and guide subsequent investigations.

\section*{Acknowledgment}

We express our gratitude to CNPq (Grant 312275/2023-4), FAPERJ (Grant E-26/204.256/2024), Kunumi, and Stone Co. for their generous support.

\bibliographystyle{IEEEtranS}
\bibliography{seaa_paper} 

\end{document}